# Comparing webometric with web-independent rankings: a case study with German universities

Mark Thamm
GESIS – Leibniz-Institute
for the Social Sciences
Lennéstr. 30, 53113 Bonn, Germany

mark.thamm@gesis.org

Philipp Mayr
GESIS – Leibniz-Institute
for the Social Sciences
Lennéstr. 30, 53113 Bonn, Germany

philipp.mayr@gesis.org

## ABSTRACT
In this paper we examine if hyperlink-based (webometric) indicators can be used to rank academic websites. Therefore we analyzed the interlinking structure of German university websites and compared our simple hyperlink-based ranking with official and web-independent rankings of universities. We found that link impact could not easily be seen as a prestige factor for universities.

## Categories and Subject Descriptors
H.5.4 [**INFORMATION INTERFACES AND PRESENTATION (I.7)**]: Hypertext/Hypermedia - *User issues*.

## 1. INTRODUCTION
Interlinking data is a fascinating new data source for the analysis of web phenomena and a feasible starting point to build a better understanding of the Web [1],[6]. In this paper we compare methodologically different approaches to rank universities. For our case study we analyzed subsets of German universities. We compare two purely webometric approaches which are based on interlinking/web data of university websites with traditional approaches which take into account e.g. judgments of students, professors and staff, publication data and other data sources .

The idea of the paper is to analyze whether there is a correlation between these different approaches to rank universities. On the long hand we try to figure out if we can use the socio-technical properties of the Web (explicated often via web links) as a proper data source for Web Science analysis.

Beside their functional role, hyperlinks can be understood as a conceptual reference to the content they are linking to. The question is: Is this kind of non-functional meaning somehow measurable?

Our assumption was:

- university websites are linking to academic and research content; so we have a significant number of non-functional links in the academic web,

- important research topics increase both, the prestige of the university as well as the number of links referring to it.

If there is a positive correlation between the prestige of an university and its inlinks, it could be seen as an effect of this non-functional nature of hyperlinks. Following the Matthew effect, more distinguished universities should earn more links than other universities.

## 2. Design of the Study
In our study we build a ranking based on the interlinking structure of the German university websites. In the following we compare this simple pure-webometric ranking with some official rankings that are based on web-independent indicators (compare with Fig.4).

### 2.1 Building a pure-webometric ranking
Search engines offer the possibility to ask for all links between two web domains [2]. Suppose a and b are websites, the search command e.g. "linkdomain:a site:b" offers the hit count estimates of all inlinks to a from b. The search command in a search engine will return results for all pages within the top level domains a and b. To obtain a number of all in- and outlinks for a list of domain names, all possible pairs of domains have to be queried. Thus we first took a list of domain names from German universities[1] to ask Yahoo for the number of links between each possible pair [2]. Therefore we used a freely available tool called LexiUrl.[2] LexiUrl automatically generates the necessary list of queries for a certain search engine from a simple list of domain names. As a result LexiUrl provides a link network graph that easily can be used with Pajek (see Fig. 3).[3] Pajek also generates a flat file, which contains all inlink amounts for each query. In addition we consult the centrality degrees closeness and hubs-authorities to build simple variants of graph based rankings. These two rankings are based on the respective centrality of each node as it can be calculated with Pajek.

Once we have all results – the hit count estimates from Yahoo as well as the network degrees from the resulting network graph - we put them into a database. Now we were able to create our own rankings, as we have counted all the links, either grouped by in- or outgoing domain name.

### 2.2 Comparison with official rankings
One of the official rankings we want to correlate with is from the Centre for Higher Education (CHE), an institute funded by the Bertelsmann Foundation.[4] CHE is a ranking of German-speaking university institutes with the primary purpose to guide freshman students in choosing the best place for their scholarly education. Its data is collected through questionnaires administered to members of departments or faculties, professors, students as well as on bibliometric analyses of the publications. CHE presents the university rankings for each discipline separately. Although CHE indicates that the data actually does not allow an overall comparison of all universities within all disciplines, we do think that comparing an overall summary of web-independent CHE data with our link based data is still useful. In order to receive a complete ranking we wrote a little program that reads the HTML from the CHE websites and transforms the colour coded judgment of each rank position in a calculable number. After we had done that for each discipline, we normalized all data with respect to the number of judgments per university.

---

[1] http://de.wikipedia.org/wiki/Liste_der_Hochschulen_in_Deutschland

[2] LexiUrl software: http://lexiurl.wlv.ac.uk/

[3] Pajekt: http://pajek.imfm.si

[4] www.che.de

With the so called Shanghai Ranking[5] we took another multi-indicator and web-independent university ranking into account. The ranking compares institutions worldwide according to a formula that includes alumni winning Nobel Prizes and Fields Medals, highly-cited researchers, articles published in famous journals and the per capita academic performance of an institution.

In addition we consult a pure webometric ranking, called Webometric Ranking of World Universities.[6] The ranking combines four indicators including the size of the pages, the number of inlinks, differentiation between filetypes and the number of papers cited by Google Scholar. Both rankings, Shanghai as well as the Webometric Ranking, are global rankings from which we took only the ranking positions of German universities. At last we simply asked Google for the general inlink amount of each domain-name (see last column of Fig.1). Again this data was put into the database.

After storing every ranking value into the database we compared each pair with the Pearson correlation coefficient.

## 3. Results

After adjusting our data of some undesirable effects (some are discussed below), we finally got 91 nodes of universities and 30,094 edges (see Fig. 3). We compared each ranking with each other using the Pearson correlation coefficient.[7]

The correlations differ in the number and selection of ranked universities. The following table indicates the amount of correlated universities (see Fig. 1. Caption below).

|    | SH | CH | IL | OL | CS | AH | GG |
|----|----|----|----|----|----|----|----|
| WT | 33 | 58 | 56 | 54 | 54 | 54 | 64 |
| SH |    | 36 | 38 | 36 | 37 | 37 | 39 |
| CH |    |    | 79 | 79 | 77 | 77 | 100 |
| IL |    |    |    |    | 91 | 91 | 91 |

**Figure 1: Total amount of correlated universities**

| **Webometrics** | WT | **Outlinks** | OL |
| **Shanghai** | SH | **Closeness** | CS |
| **CHE** | CH | **Authorities &Hubs** | AH |
| **Inlinks** | IL | **Google** | GG |

---

[5] http://www.arwu.org/ARWU2010.jsp

[6] http://www.webometrics.info

[7] We also tried Spearman's correlation coefficient with a similar result.

The Pearson correlation coefficient is shown in the table below (see Fig. 2).

|    | SH | CH | IL | OL | CS | AH | GG |
|----|----|----|----|----|----|----|----|
| WT | 0.55 | 0.04 | 0.12 | -0.07 | -0.16 | 0.17 | 0.17 |
| SH |    | 0.11 | -0.03 | -0.08 | -0.22 | -0.07 | 0.16 |
| CH |    |    | 0.01 | 0.06 | -0.13 | 0.04 | 0.05 |
| IL |    |    |    |    | 0.84 | 0.88 | -0.04 |

**Figure 2: Correlation of university rankings (Pearson correlation)**

There are only three noteworthy correlations. First of all, that is the correlation between the Webometrics and the Shanghai ranking (p=0.55). Furthermore there are two strong positive correlations between the centrality degrees - closeness (p=0.84) and the combined hubs-authorities (p=0.88) value - and the in- and outlink amount. Since the two ranges were taken from the same graph (see Fig. 3), they must have a good correlation, as the number of in- and outlinks and centrality measures are not independent of each other. These correlations, however, say little about the relationship of the graph with the other rankings and will therefore not be further interpreted. All other measurement parameters are close to zero or negativ.

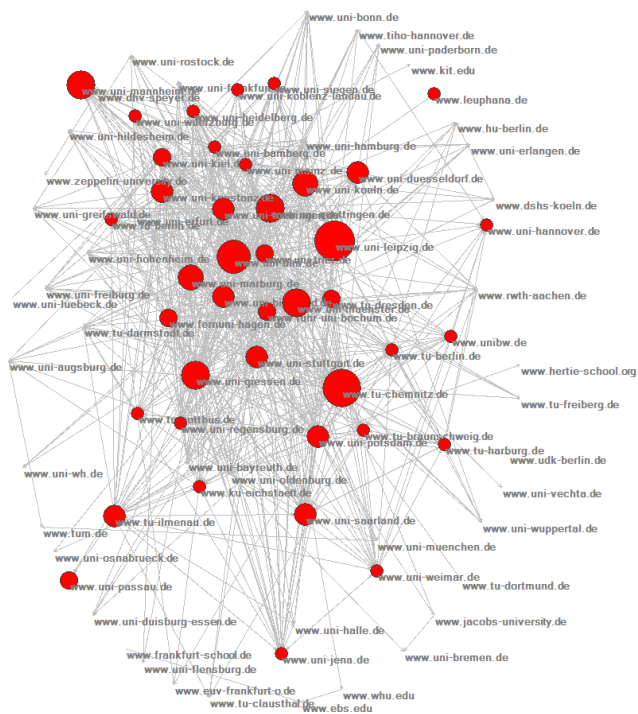

**Figure 3: Adjusted interlinking graph of university websites in Germany with 30,094 links from finally 91 universities. The size of nodes depends on the number of websites per domain.**

## 4. Discussion

Apart from the rankings Webometrics and Shanghai (p=0.55) there are no obvious correlations in our data. That means in general, hyperlinks cannot easily be used as ranking indicators for academic website in this simplistic way.. Obviously the objective of hyperlinking is more complex. Deriving indicators directly from the Web is not that simple due to a multitude of motivations to create hyperlinks. Although we have adjusted our data from such obvious cases, we are not sure whether there exist some more.

When preparing the data, we quickly came across a general problem of link analysis. A certain set of significant highly linked university pages owes this position due to a purely technical aspect. E.g. the university website of Kassel was in a top position because it was referenced by the website of Marburg, Giessen and Mainz due to technical links from a content management system. All stylesheets, scripts, images are referencing the domain name of Kassel. Those links are counted although the motivation of creation was obviously not a conceptual reference to the content of Kassel.

Does that mean hyperlinks do not have a conceptual meaning as it is assumed above? In our opinion, hyperlinks cannot be understood simply as a measure of appreciation of the linked resource. The reasons for the existence of a hyperlink are too multifunctional [3]. However, we still think that hyperlinks represent prestige in web infrastructures but in our restricted case study with German university website, we cannot observe these prestige differences expressed via simple link counts. In addition hyperlinks need to be classified, at least to distinguish between pure functional and substantive meaning. There have already been made some efforts on general link classification. Thelwall [3] and Stuart [4] for example are using a distinction made on the objectives of a link to introduce a general typology of hyperlinks. This approach might have produced a better correlation between the number of meaningful inlinks and a ranking position established by any web-independent method.

Basically, there exist fundamental problems in the underlying services provided by public search engines [7]. The reason why we have used Yahoo instead of Google is that Google has stopped its service in late 2010. Since April 2011, Yahoo has also prevented the ability to query linking amounts via the search engine.

Another problem in our approach is the lack of control over the interlinking data. In order to classify hyperlinks, Web Science researchers need to have access to the raw data coming from focused crawls. As our example with the technical links to the Kassel university website shows, it should be possible to roughly assess from which part of an HTML page a link is received from. In order to avoid those problems and to be able to classify hyperlinks before counting in the future, we will replace LexiURL with a complete crawl of the university websites.

|  | CHE | Shanghai | Webometrics | Inlinks |
|---|---|---|---|---|
| 1 | www.uni-konstanz.de | www.uni-muenchen.de | www.ruhr-uni-bochum.de | www.uni-leipzig.de |
| 2 | www.uni-freiburg.de | www.tum.de | www.uni-marburg.de | www.ruhr-uni-bochum.de |
| 3 | www.uni-jena.de | www.uni-heidelberg.de | www.uni-konstanz.de | www.uni-muenster.de |
| 4 | www.uni-heidelberg.de | www.uni-goettingen.de | www.uni-giessen.de | www.uni-marburg.de |
| 5 | www.uni-muenster.de | www.uni-frankfurt.de | www.uni-jena.de | www.uni-koeln.de |
| 6 | www.uni-marburg.de | www.uni-freiburg.de | www.uni-wuerzburg.de | www.uni-goettingen.de |
| 7 | www.uni-greifswald.de | www.uni-wuerzburg.de | www.uni-ulm.de | www.uni-giessen.de |
| 8 | www.uni-goettingen.de | www.uni-mainz.de | www.uni-halle.de | www.uni-bielefeld.de |
| 9 | www.uni-muenchen.de | www.uni-muenster.de | www.uni-bayreuth.de | www.uni-konstanz.de |
| 10 | www.uni-kiel.de | www.uni-koeln.de | www.uni-greifswald.de | www.tu-berlin.de |

**Figure 4: Top 10 ranking from CHE, Shanghai, Webometrics and Inlinks amount**